\documentclass[referee]{raa}           
\usepackage{graphicx,times}
\usepackage{natbib}
\usepackage{amssymb,amsmath}
\bibpunct{(}{)}{;}{a}{}{,}
\usepackage{multirow}
\usepackage{longtable}
\usepackage{graphicx}
\usepackage{CJKutf8}

\usepackage[pagebackref=true]{hyperref}

\usepackage{epstopdf}
\usepackage{natbib}
\usepackage{graphicx}
\usepackage{CJK}

\usepackage[percent]{overpic}
\usepackage{times}
\usepackage{graphicx}
\usepackage{amsmath}
\usepackage{natbib}
\usepackage{txfonts}
\usepackage{xcolor}
\usepackage[utf8]{inputenc}
\usepackage[overload]{empheq}
\usepackage{verbatim}

\newcommand{\ks}{$K_{\rm S}$}

\newcommand{\gbp}{$G_{\rm BP}$}
\newcommand{\grp}{$G_{\rm RP}$}

\newcommand{\ebr}{$E(G_{\rm BP}-G_{\rm RP})$}

\newcommand{\ebv}{$E(B-V)$}

\newcommand{\mg}{$M_{G}$}

\newcommand{\teff}{$T_{\rm eff}$}
\newcommand{\feh}{$\rm [Fe/H]$}

\begin{document}
\begin{CJK*}{UTF8}{gbsn}

   \title{Stellar parameters, extinction, and distances for stars in SMSS DR2 by SPar method}

 \author{Mingxu Sun
    \inst{1}
    \and
    Bingqiu Chen
    \inst{2}
    \and
    Baokun Sun
    \inst{2}
    \and
    Tao Wang
    \inst{2}
    \and
    Zheng Yu
    \inst{2}
    \and
    Baisong Zhang
    \inst{2}
    \and
    Lin Zhang
    \inst{2}
    \and
    Yuxi Bao
    \inst{2}
    \and
    Guangya Zeng
    \inst{2}
    \and
    Ming Yang
    \inst{3}
    \and
    Wenyuan Cui
    \inst{1}
    }
    
    \institute{$^{1}${Department of Physics, Hebei Key Laboratory of Photophysics Research and Application, Hebei Normal University, Shijiazhuang 050024, People's Republic of China}\\
    $^{2}${South-Western Institute for Astronomy Research, Yunnan University, Kunming 650500, People's Republic of China; {\it bchen@ynu.edu.cn}}\\
    $^{3}${Key Laboratory of Space Astronomy and Technology, National Astronomical Observatories, Chinese Academy of Sciences, Beijing 100101, People's Republic of China}\\
    {\small Received 10 November 2024; accepted 24 March 2025}
    }

\abstract{
The availability of large datasets containing stellar parameters, distances, and extinctions for stars in the Milky Way, particularly within the Galactic disk, is essential for advancing our understanding of the Galaxy's stellar populations, structure, kinematics, and chemical evolution. In this study, we present a catalog of stellar parameters, including effective temperature (\teff), metallicity (\feh), absolute magnitudes ($M_{G}$), distances ($d$), and reddening values (\ebr), for a sample of 141 million stars from the SkyMapper Southern Survey (SMSS). These parameters are derived using the SPar algorithm, which employs a fitting procedure to match multi-band photometric observations and estimate the stellar properties (\teff, \feh, $M_G$, $d$, and \ebr) on an individual star basis, following the methodology outlined in our previous work. This study successfully determines stellar parameters and extinction values simultaneously for stars located in high and low Galactic latitudes. The resulting stellar parameters and extinction values are in good agreement with results from other studies, demonstrating the robustness of our method. We derive a temperature dispersion of 195\,K and a metallicity dispersion of 0.31\,dex when comparing our results with spectroscopic data. The catalog produced in this work provides a valuable resource for future research on the Galactic metallicity distribution function, the structure of the Galaxy, three-dimensional extinction mapping, and other related astrophysical topics.
}

   \authorrunning{Sun et al. }            
   \titlerunning{Stellar parameters, extinction, and distances for stars in SMSS DR2 by SPar method}  
   \maketitle

%

\section{Introduction} \label{sec:intro}

Understanding the formation and evolution of the Milky Way requires precise measurements of stellar atmospheric parameters, distances, and extinction for large samples of stars, particularly those in the Galactic disk. Such measurements allow us to characterize the properties of stellar populations and their spatial distributions across the Galaxy (\citealt{2011A&A...530A.138C}; \citealt{2013MNRAS.434.3165P}; \citealt{2016ARA&A..54..529B}; \citealt{2021ApJ...912..147W}; \citealt{Gaia2018}).

Large-scale spectroscopic surveys have revolutionized the field by providing precise stellar parameters for tens of millions of stars. Notable examples include the Sloan Extension for Galactic Understanding and Exploration (SEGUE; \citealt{yanny2009segue}), the Large Sky Area Multi-Object Fiber Spectroscopic Telescope (LAMOST; \citealt{Luo2015}), the Apache Point Observatory Galactic Evolution Experiment (APOGEE; \citealt{SDSSDR17}), the Galactic Archaeology with HERMES (GALAH; \citealt{Buder2021}), and the Dark Energy Spectroscopic Instrument (DESI; \citealt{2022AJ....164..207D}). These surveys have provided high-quality datasets with broad coverage of parameter space, enabling the inference of stellar properties from photometric data. Complementing spectroscopic surveys, narrow- and medium-band photometric surveys, such as the SkyMapper Southern Survey (SMSS; \citealt{Wolf2018, Onken2019}), the Javalambre/Southern Photometric Local Universe Survey (J/S-PLUS; \citealt{Cenarro2019}), and the Sloan Digital Sky Survey (SDSS; \citealt{SDSS2000}), have demonstrated the ability to estimate stellar parameters with high accuracy through their carefully designed filter sets.

Recent studies have demonstrated the efficacy of photometric methods for deriving stellar parameters. For example, \citet{yuan2015a,yuan2015b} used empirical metallicity-dependent stellar loci to estimate photometric metallicity for 500,000 FGK dwarf stars in Stripe 82. \citet{Chen2019} derived intrinsic colors and extinction values for 23 million stars in the Galactic disk using optical and near-IR photometry, with spectroscopic data from LAMOST and SDSS serving as training samples. Extending the Yuan et al. methods to red giants, \citet{Zhang2021} developed metallicity-dependent stellar loci to estimate metallicities from SDSS photometry, enabling the identification of metal-poor red giants. Leveraging the sensitivity of the SMSS \(uvgriz\) filters, \citet{huang2019} constructed empirical relationships between photometric colors and stellar parameters, deriving accurate atmospheric parameters for approximately one million red giants. Building on this, \citet{Huang2022} estimated stellar parameters for 24 million stars using SMSS photometry, and \citet{Chiti2021} utilized a grid-based synthetic photometry approach to derive metallicities for over 250,000 stars from SMSS data. Other notable efforts include \citet{Xu2022}, who estimated metallicities for 27 million stars using Gaia EDR3 photometry with LAMOST spectroscopic data as training samples, and \citet{Yang2022}, who derived stellar parameters from J-PLUS photometry \citep{Cenarro2019}.

For stars at high Galactic latitudes, extinction effects are minimal, allowing for precise parameter estimates using two-dimensional extinction maps (e.g., \citealt{1998ApJ...500..525S}). However, in the low-latitude Galactic disk, where dust extinction is significant, simultaneous measurements of extinction and metallicity are essential to reduce systematic errors in parameter estimation \citep{Andrae2022, 2023AJ....166..126S}. This highlights the need for integrated approaches that can account for extinction effects while deriving stellar parameters.

In this study, we employ the SPar algorithm \citep{2023AJ....166..126S} to estimate stellar atmospheric parameters, extinction, and distances for a sample of 140 million stars from SMSS DR2. The SPar algorithm relies on empirical stellar libraries trained on LAMOST spectroscopic data \citep{Luo2015}. Our dataset combines $uvgriz$ photometry from SMSS with $JH$\ks\ measurements from the Two Micron All Sky Survey (2MASS; \citealt{Skrutskie2006}), $W1W2$ photometry from the Wide-field Infrared Survey Explorer (WISE; \citealt{WISE2010}), and $G$\gbp\grp\ data from Gaia DR3 \citep{Gaiadr3}. By matching observations to the stellar template library, we derive key stellar parameters, including effective temperature (\teff), metallicity (\feh), absolute magnitude (\mg), distance (\textit{d}), and extinction (\ebr) for the individual stars.

Unlike previous algorithms such as Star-Horse \citep{Anders2019, Anders2020} and GSP-Phot \citep{Andrae2022}, which rely on theoretical stellar models that may introduce systematic uncertainties \citep{Green2021}, our approach is based on empirical relationships derived directly from observational data. By leveraging the sensitivity of the SMSS \(uv\) filters to metallicity and utilizing a robust training dataset, we achieve accurate estimates of stellar parameters and extinction. Compared to prior studies using SMSS data, our work stands out for its larger sample size and the simultaneous estimation of stellar parameters and extinction, enabling precise characterization of stars in the Galactic disk and improving our understanding of the Galactic disk’s structure and evolution.

\section{Data}\label{sec:dat}

In this study, we utilize the $uvgriz$ photometric data from the SMSS DR2 \citep{Onken2019, Wolf2018}. SMSS is a hemispheric survey conducted at Siding Spring Observatory in Australia using the SkyMapper Telescope. The catalog contains a total of 505 million sources. The survey reaches depths of 19.7 to 21.7\,mag across its six bands: $u$, $v$, $g$, $r$, $i$, and $z$. Compared to the single $u$ band in SDSS, characterized by ($\lambda_{\rm cen}$/FWHM) = (358 nm/55 nm), SMSS uses two separate filters in the blue: a violet $v$ band (384 nm/28 nm) and a more ultraviolet $u$ band (349 nm/42 nm). This configuration enhances the sensitivity to variations in metallicity. The internal photometric precision of SMSS DR2 is 1\% for the $u$ and $v$ bands and 0.7\% for the $griz$ bands. Over 21,000 square degrees are covered in certain filters, with more than 7,000 square degrees surveyed deeply across all six bands. 

\begin{figure*}
\centering
\includegraphics[width=0.8\textwidth]{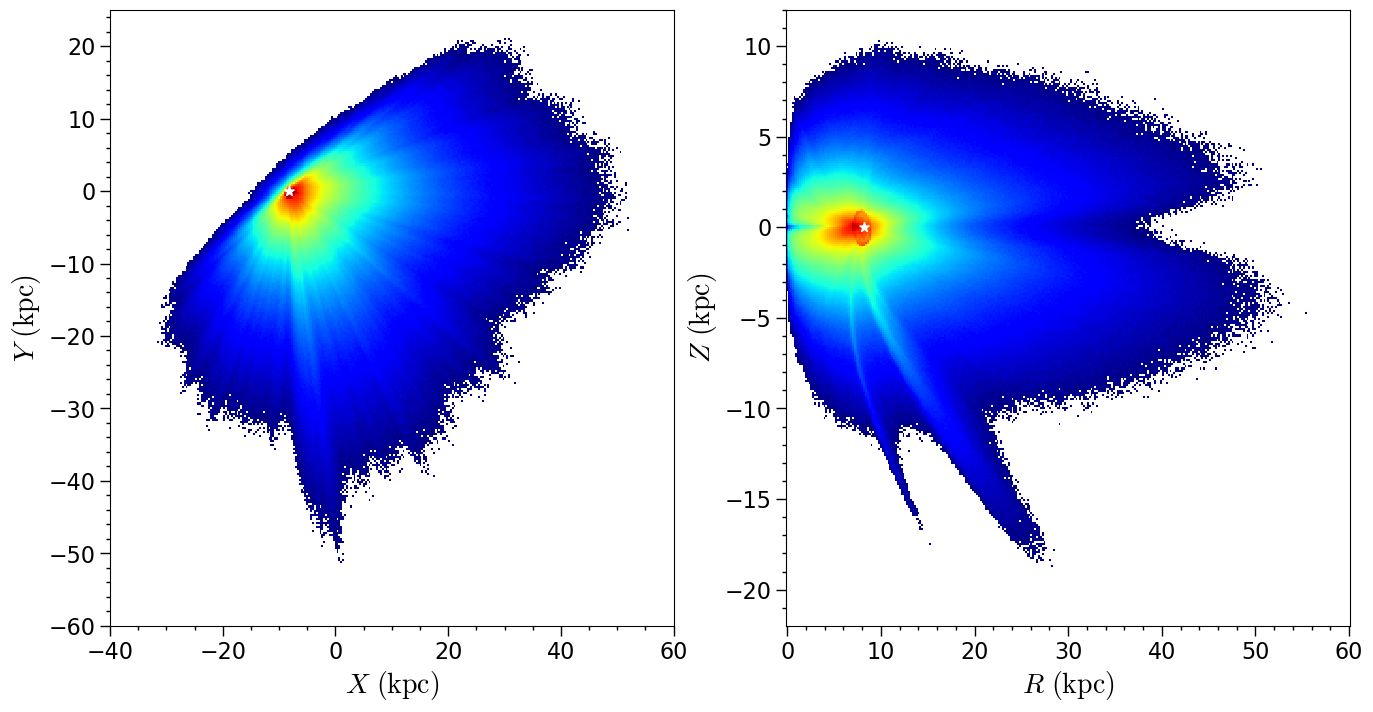}
\caption{The number-density distribution of all stars in SMSS in the Galactic plane ($X$-$Y$, left panel) and perpendicular to it ($R$-$Z$, right panel). The Sun's position is marked with white stars, located at ($X,~Y$) = ($-$8.178, 0)\,kpc in the left panel and at ($R,~Z$) = (8.178,~0)\,kpc in the right panel.}
\label{xydc}
\end{figure*}

In Fig.~\ref{xydc}, we present the spatial distribution of stars in SMSS within the Galactic Cartesian coordinate system. The $X$-axis points toward the Galactic center, the $Y$-axis toward the direction of Galactic rotation, and the $Z$-axis toward the north Galactic pole. The XYZ values for individual stars are calculated using their Galactic coordinates (l, b) and distances (d), where distances are simply derived as the inverse of Gaia DR3 parallaxes (d = $1/\varpi$). Stars without Gaia parallaxes or with negative parallaxes are excluded, accounting for approximately 22\% of the total sample. Notably, distance uncertainties increase significantly for stars at greater distances due to the large errors in Gaia parallax measurements. This distribution clearly traces the disk structure of the Milky Way.

To enhance the dataset, we combine SMSS data with infrared photometry from the 2MASS and WISE catalogs. The 2MASS survey, conducted across the entire sky, uses three filters: $J$, $H$, and \ks\ \citep{Skrutskie1997}. Its source catalog contains 470 million objects. The WISE survey, using the $W1$, $W2$, $W3$, and $W4$ bands, provides additional infrared data. For this study, we use the AllWISE Source Catalog \citep{Kirkpatrick2014}, which contains 748 million sources. To ensure high data quality, we limit our analysis to the $W1$ and $W2$ bands due to their superior sensitivity and angular resolution compared to $W3$ and $W4$. Furthermore, we incorporate Gaia DR3 data \citep{Gaiadr3}, which provides essential astrometric parameters, including positions, parallaxes, and proper motions, as well as photometry in the $G$, \gbp, and \grp\ bands. Gaia DR3 encompasses 1.8 billion sources, making it an invaluable resource for cross-matching and parameter estimation.

\begin{figure*}
    \centering
    \includegraphics[width=0.8\textwidth]{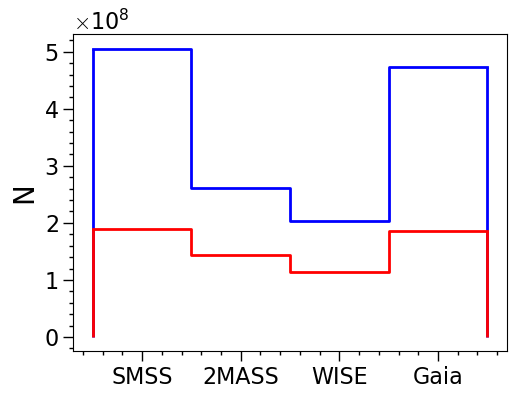}
    \caption{Histogram illustrating the contributions of different catalogs (blue line) to the merged catalog (red line). Most SMSS sources have corresponding data from Gaia, while approximately half have matches in 2MASS and WISE.}
    \label{distry}
\end{figure*}

The SMSS data is cross-matched with 2MASS, AllWISE, and Gaia using a matching radius of 1.5\,arcsec, resulting in the creation of the SMSS-2MASS-AllWISE-Gaia (SSAG) catalog. To ensure data quality, we apply the following selection criteria: (1) sources must have measurements in the SMSS $gri$ bands and at least one additional band, and (2) photometric errors in the SMSS $gri$ bands and at least one additional band must be less than 0.1\,mag. After applying these criteria, the SSAG catalog includes a total of 188,464,828 sources.

Fig.~\ref{distry} presents a histogram illustrating the contributions of different catalogs to the merged dataset. Our selection criteria, which require all SMSS sources to have complete observations in the $gri$ bands and photometric errors below 0.1\,mag in these bands, exclude a large portion of the original SMSS sources. Despite this, the majority of SMSS sources in the final sample are successfully matched with Gaia data, and over half of the sources are also cross-matched with 2MASS and WISE.

\begin{figure*}
    \centering
    \includegraphics[width=0.8\textwidth]{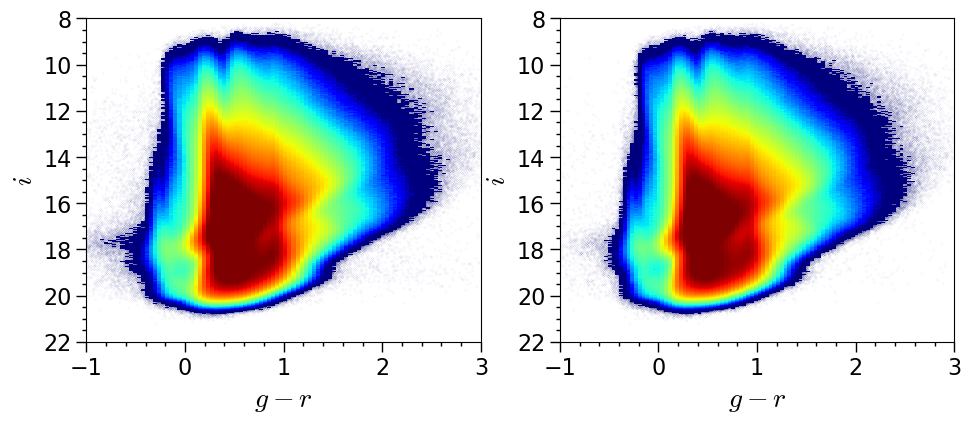}
    \caption{Color-magnitude diagrams $(g-r)$ vs. $i$ of the original SMSS catalog (left) and the SSAG catalog (right). The diagrams show no significant differences between the original and merged catalogs.}
    \label{colma}
\end{figure*}

Fig.~\ref{colma} displays the color-magnitude diagrams ($g-r$ vs. $i$) for the original SMSS catalog and the merged SSAG catalog. The overall distributions in both diagrams show no significant differences, indicating that the merging process preserves the consistency and integrity of the data. The $g-r$ values primarily range from $-$2.5 to 2.5\,mag, while the $i$ magnitudes span from 9 to 21\,mag. This consistency suggests that, while the sample size is reduced due to the selection criteria, the proportion of different stellar populations within the sample remains unchanged.

\section{Method}\label{sec:emp}

The methodology employed in this study is based on the SPar algorithm described by \citet{2023AJ....166..126S}. This algorithm establishes relationships between stellar parameters—effective temperature (\teff), metallicity (\feh), and absolute magnitude in the $G$ band (\mg)—and the absolute magnitudes across all filters using empirical stellar templates. For each star in the SSAG catalog, we perform a minimum $\chi^2$ fitting of the multiband photometry and parallax data to derive initial parameter estimates. These estimates are then refined using a Markov Chain Monte Carlo (MCMC) analysis, which provides the final stellar parameters along with their uncertainties. Below, we provide a concise summary of the methodology, while further details can be found in \citet{2023AJ....166..126S}:

\begin{enumerate}
    \item We construct an empirical stellar library by selecting a sample of stars from LAMOST and Gaia DR3 that have well-determined atmospheric parameters, distances, and extinction values. This sample is supplemented with optical and near-infrared photometry from SMSS, 2MASS, and WISE. The library includes stellar parameters and the corresponding absolute magnitudes for each filter.
    \item A Random Forest Regression algorithm is applied to the stellar library to map the relationships between the stellar parameters (\teff, \feh, and \mg) and the absolute magnitudes in each passband. This process generates empirical stellar templates used for parameter estimation.
    \item Distance estimation is performed using observed magnitudes in optical bands and the empirical stellar templates. For a given star, the stellar templates, combined with assumed values of \teff, \feh, \mg, and \ebr, are used to predict the `distance modulus-corrected' magnitudes $M'_x$ in the $x$ filter, using the relation:  $M'_x = M_x$(\teff, \feh, \mg) + $A_x$(\ebr),  where $A_x$(\ebr) represents the extinction in filter $x$. The distance modulus $\mu$ is then derived by subtracting $M'_x$ from the observed magnitudes $m_x$: $\mu = m_x - M'_x$. This calculation is restricted to optical filters, specifically Gaia $G$, \gbp, \grp, and SMSS $gri$. This is due to that the Gaia magnitudes have high photometric accuracies of and SMSS $gri$ bands are available for all sources in the final merged catalog. The standard magnitude relation, $m_x = M_x + A_x + \mu$, is then used to simulate the apparent magnitudes in the individual passbands.
    \item The observed magnitudes across all available filters and the observed parallaxes are modeled as a function of four free parameters: \teff, \feh, \mg, and \ebr. Initially, a coarse grid search is performed using a minimum $\chi^2$ method to identify the best-fit values of \teff, \feh, \mg, and \ebr\ for each star in the SSAG catalog.
    \item Stars with acceptable fits, defined as those with $\chi^2 < 10$, are selected. The best-fit parameters from the $\chi^2$ minimization serve as initial values for an MCMC analysis. This step refines the parameter estimates and provides their final values along with associated uncertainties.
\end{enumerate}

\section{Results and Discussion}

In this section, we present the derived stellar parameters for our sample, including effective temperatures (\teff), metallicities (\feh), absolute magnitudes (\mg), reddening values (\ebr), and distances ($d$). We also analyze and discuss the overall distribution of these parameters, as well as the spatial distribution of the stars within the sample.

\begin{figure*}
\centering
\includegraphics[width=1\textwidth]{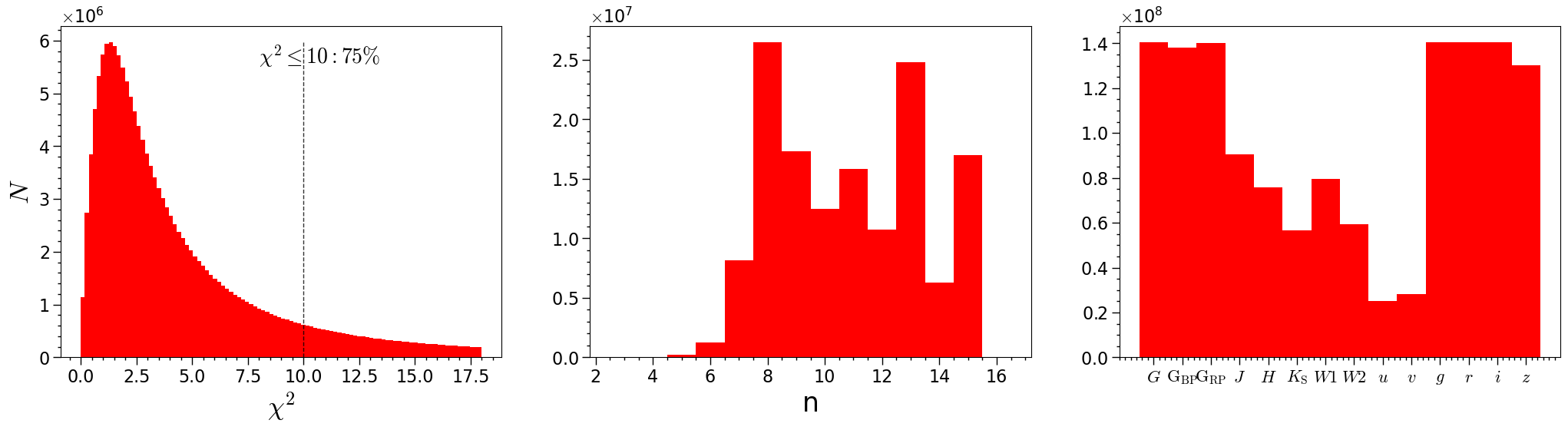}
\caption{Left: The distribution of the $\chi^2$ values with a threshold of $\chi^2 = 10$. Middle: The distribution of $n$ (the number of adopted measurements, including multi-band photometry and Gaia parallax) for stars with $\chi^2 < 10$. Right: The distribution of photometric bands for stars with $\chi^2 < 10$.}
\label{distr}
\end{figure*}

Fig.~\ref{distr} illustrates the observational data used in this study, including multi-band photometry and Gaia parallax measurements (when available), along with the $\chi^2$ distribution for all stars in the SSAG sample. As expected, the $\chi^2$ distribution peaks near 2 and exhibits an extended tail. Of the 188,464,828 stars in the SSAG catalog, approximately 75\% (140,599,779 stars) have $\chi^2$ values below 10, which we use as a threshold for selecting high-quality fits. The stellar parameters for these stars were refined using the MCMC analysis. For the sources with $\chi^2 < 10$, the majority have more than eight independent measurements, combining magnitudes from different photometric filters and Gaia parallax. Nearly all stars in this subset have observations in the SMSS $gri$ bands and Gaia $G$, \gbp, and \grp\ bands. Additionally, approximately 25 million stars (18\%) have observations in the SMSS $u$ band, while 28 million stars (20\%) have data in the SMSS $v$ band. This broad coverage across multiple filters enables robust parameter estimation. A detailed table containing the derived parameters for the 141 million stars with $\chi^2 < 10$ is available on the website at \url{https://nadc.china-vo.org/res/r101372/}. This publicly accessible dataset provides a valuable resource for further studies of stellar populations and Galactic structure.

\begin{figure*}
\centering
\includegraphics[width=0.8\textwidth]{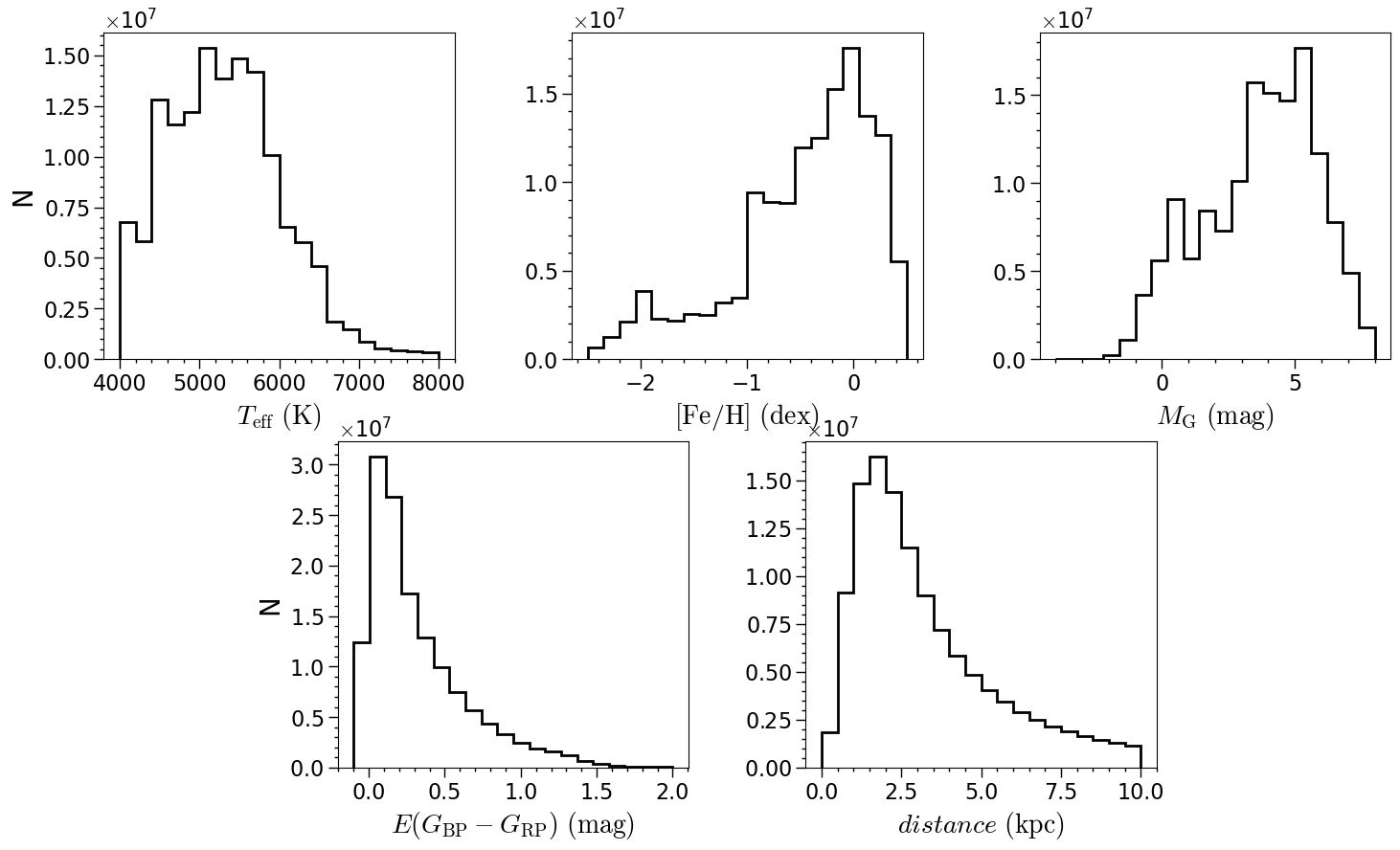}
\caption{Distributions of the derived parameters for all stars in our final catalog: \teff\ (top-left), \feh\ (top-middle), \mg\ (top-right), \ebr\ (bottom-left), and distance (bottom-right).}
\label{csfb}
\end{figure*}

Fig.~\ref{csfb} illustrates the parameter distributions for the 141 million stars in our catalog, including effective temperature (\teff), metallicity (\feh), absolute magnitude (\mg), extinction (\ebr), and distance. The majority of stars have \teff\ values between 5000 and 6000\,K, reflecting the dominance of mid-type stars in the sample. Metallicity values (\feh) are primarily concentrated in the range $-$0.4 to 0.2\,dex, and \mg\ values are clustered between 3 and 6\,mag. For the reddening values (\ebr), most stars exhibit values between 0 and 0.3\,mag. Distances are predominantly within the range of 1000 to 3000\,pc.

To emphasize the uniqueness of our stellar sample, Table~\ref{com} provides a comparative overview of the stellar parameter ranges from this study and those from \citet{Huang2022}, \citet{Xu2022}, and \citet{2023MNRAS.524.1855Z}. Compared to \citet{Huang2022} and \citet{Xu2022}, our work has a significantly larger sample size and benefits from the simultaneous estimation of stellar parameters and extinction, which allows for more precise characterization of stellar properties, particularly in the Galactic disk. Unlike the sample in \citet{2023MNRAS.524.1855Z}, which includes stars from brighter magnitude ranges, our catalog extends to much fainter stars, providing a complementary dataset for probing stellar populations at further distances.

\begin{table}
\centering
\caption{Comparison of Stellar Sample Properties}
\label{com}
\begin{tabular}{lcccc}
\hline
Parameter & This Work & H22 & Xu22 & Z23 \\ 
\hline
$G$-band Range (mag)   & 12.4--20.1 & 11.0--18.1 & 10.9--16.0 & 11.7--17.6 \\
Survey Data Used        & SMSS, 2MASS, WISE, Gaia & SMSS, Gaia & Gaia & Gaia XP, 2MASS, WISE \\
Galactic Latitude ($|b|$) Range & 0°--90° & 10°--90° & 10°--90° & 0°--90° \\
Number of Stars         & 140 million & 24 million & 26 million & 220 million \\
Giant-to-Dwarf Ratio    & 29\% & 28\% & 32\% & 29\% \\
Stellar Types           & AFGK & FGK & FGK & BAFGKM \\
Metallicity Range (dex) & $-$2.50--0.50 & $-$3.45--0.64 & $-$3.76--1.00 & $-$4.0--1.0 \\
\hline
\end{tabular}
\end{table}

In this section, we will discuss the precision of the resultant effective temperature (\teff), metallicity (\feh), extinction (\ebr), and distance for the stars in our catalog. To evaluate the accuracy of our parameter estimates, we compare our results with those from previous studies, including large-scale spectral surveys. The parameter distributions (\feh\ and \ebr) are examined spatially across the Milky Way to identify trends and correlations, providing a foundation for further exploration of stellar populations and Galactic structure. We also compare our findings with related studies to highlight the scientific implications of our work and identify potential avenues for future research.

\subsection{Discussion of \teff\ and \feh}
\label{subsec:comparison_teff_feh}

In this section, we compare the derived effective temperatures (\teff) and metallicities (\feh) with results from multiple spectroscopic surveys, including LAMOST DR11 (L11; \citealt{Luo2015}), GALAH DR3 (G3; \citealt{Buder2021}), and APOGEE DR17 (A17; \citealt{SDSSDR17}). The training sample used in this study is based on LAMOST DR8, which partially overlaps with L11 but remains independent from G3 and A17. These surveys are treated as benchmarks to assess the accuracy of our results. Additionally, we include three more catalogs for comparision: Gaia23 \citep{2023A&A...674A...1G}, Z23 \citep{2023MNRAS.524.1855Z}, and H22 \citep{Huang2022}. Cross-matching was performed between the selected catalogs and spectroscopic surveys. We use the common sources shared between these catalogs and the spectroscopic surveys for comparison.

\begin{figure*}
    \centering
    \includegraphics[width=0.8\textwidth]{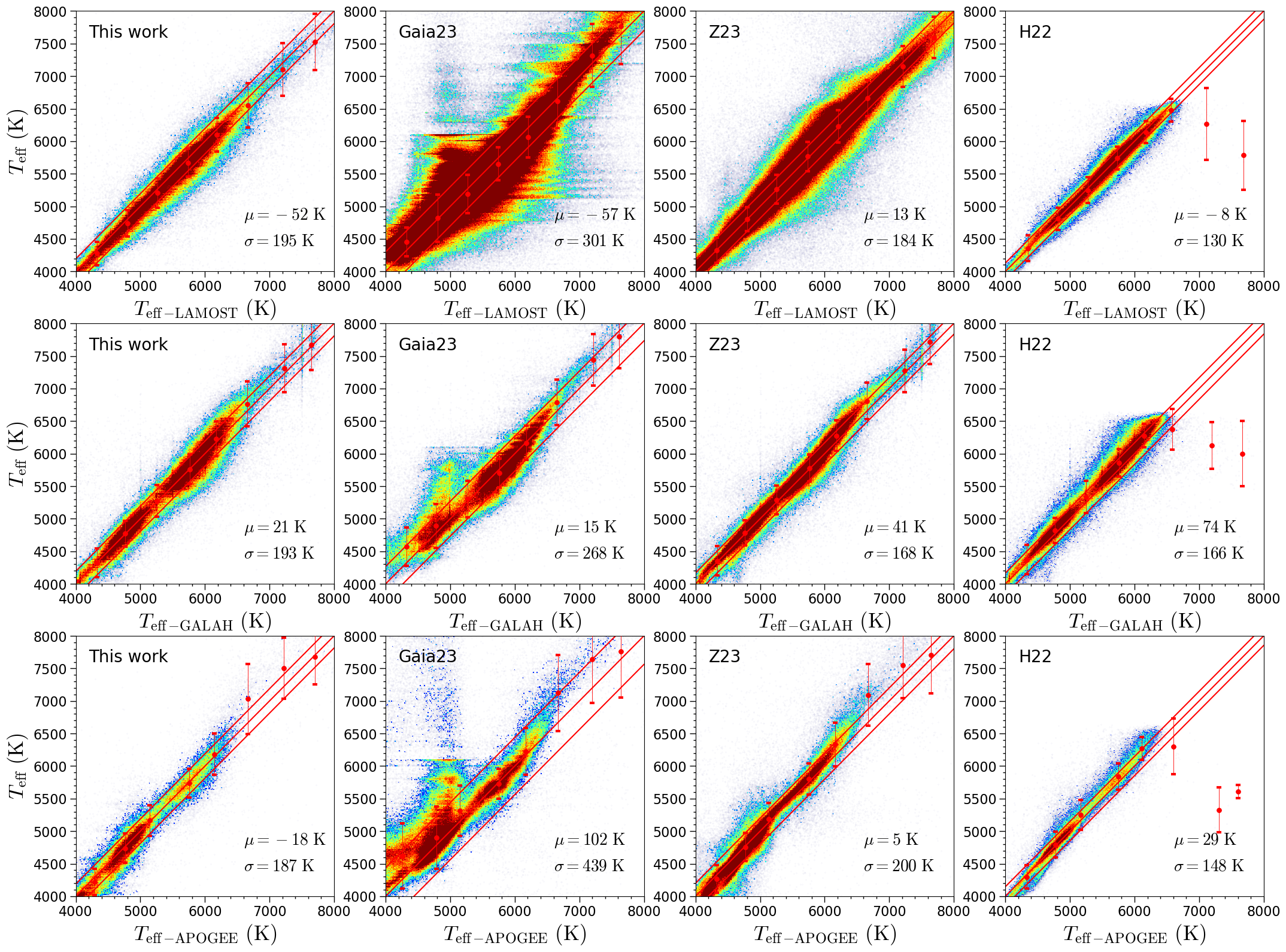}
    \caption{Comparison of \teff\ values from this study (first column), Gaia23 (second column), Z23 (third column), and H22 (fourth column) with those from LAMOST (top panels), GALAH (middle panels), and APOGEE (bottom panels). Red dots with error bars represent the average and dispersion values. Red lines indicate the line of equality and the line of equality plus/minus the dispersion of the overall sample.}
    \label{xtlag}
\end{figure*}

\begin{figure*}
    \centering
    \includegraphics[width=0.8\textwidth]{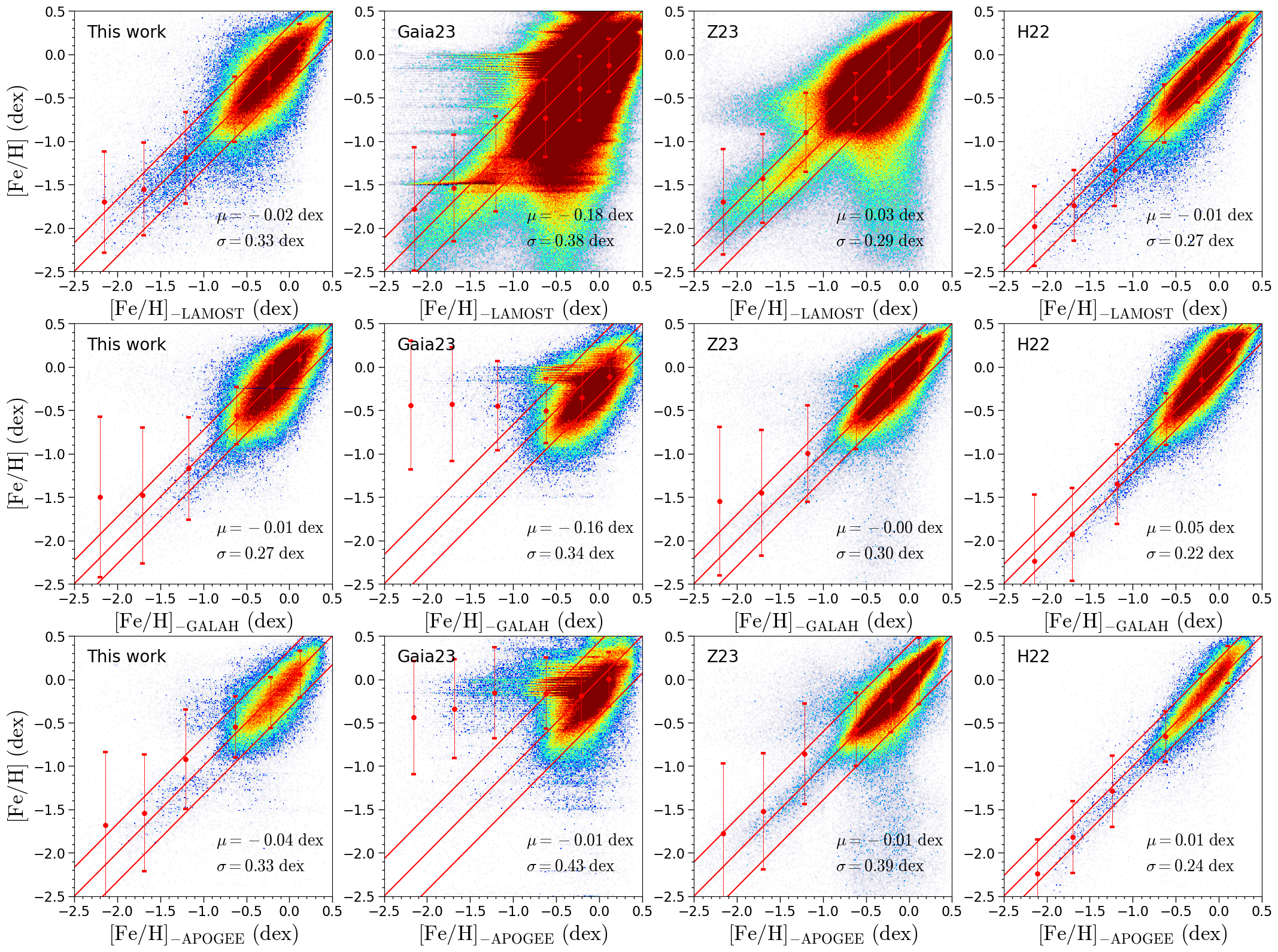}
    \caption{Similar to Fig.~\ref{xtlag}, but for \feh.}
    \label{xtlagf}
\end{figure*}

\begin{figure*}
    \centering
    \includegraphics[width=0.5\textwidth]{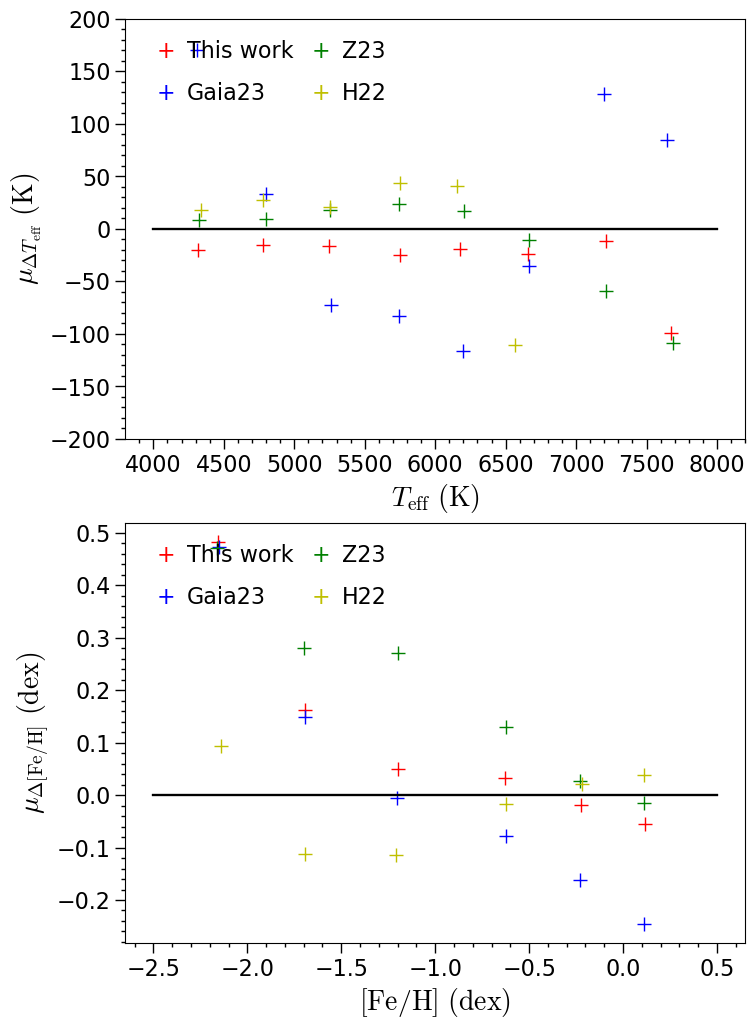}
    \caption{Median differences between this study, Gaia23, Z23, and H22 compared to spectroscopic surveys for \teff\ (top panel) and \feh\ (bottom panel). Different colors represent results from different works.}
    \label{xtlagqm}
\end{figure*}

Fig.~\ref{xtlag} compares the \teff\ values derived in this study with those from the spectroscopic surveys. Our results show good consistency with L11, G3, and A17, with median differences ($\mu$) and standard deviations ($\sigma$) of $\mu = -52$\,K and $\sigma = 195$\,K for L11, $\mu = 21$\,K and $\sigma = 193$\,K for G3, and $\mu = -18$\,K and $\sigma = 187$\,K for A17. The dispersion in differences increases with effective temperature. For G3, minor vertical stripes are visible in the comparison, which are also present when comparing Gaia23, Z23, and H22 with G3. This could indicate systematic effects related to the G3 observation template. Additionally, G3 and A17 exhibit two distinct parameter clustering regions, likely due to selection effects in these surveys. When compared with Gaia23, Z23, and H22, our \teff\ results have accuracy comparable to Z23 and H22, and slightly better than Gaia23. Notably, Gaia23 shows a systematic overestimation of \teff\ around 5000\,K and at higher temperatures compared to the spectroscopic surveys. In particular, Gaia23 exhibits distinct horizontal stripes in comparisons with L11, G3, and A17. Above 6600\,K, deviations in Z23 and H22 become apparent, primarily due to a few outlier points. H22, for instance, only includes stars with \teff\ below 6600\,K, which limits comparisons in this range.

For \feh, our results also exhibit good agreement with the spectroscopic surveys, as shown in Fig.~\ref{xtlagf}. The median differences and standard deviations are $\mu = -0.02$\,dex and $\sigma = 0.33$\,dex for L11, $\mu = -0.01$\,dex and $\sigma = 0.27$\,dex for G3, and $\mu = -0.04$\,dex and $\sigma = 0.33$\,dex for A17. The dispersion increases at lower metallicities, particularly for \feh\ $< -1$, where the differences become more pronounced. For \feh\ $< -2$, the differences are even larger, though this region contains relatively few sources. Our \feh\ results have accuracy comparable to those of Z23 and H22 and better than Gaia23. In the metallicity range of $-0.5$ to $0$\,dex, Gaia23 tends to underestimate values compared to spectroscopic surveys, whereas our results, Z23 and H22 show better agreement. Gaia23 also exhibits horizontal stripes in the \feh\ comparisons, similar to those seen in the \teff\ results. In the comparison plots of G23 and Z23, subsets of stars exhibit significant discrepancies in [Fe/H] when compared to spectroscopic results. However, such discrepancies are absent in our results and those of H22. Upon further examination, we identified that the stars exhibiting significant deviations in metallicity in the G23 and Z23 studies are primarily disk dwarf stars. Upon further examination, we found that the stars exhibiting significant deviations in metallicity in the G23 and Z23 studies are primarily disk dwarf stars. However, they account for only about 1\% of the dwarf star sample in the common dataset between these studies and the spectroscopic surveys. No significant trends in distance or extinction were identified among them. These deviations are likely statistical in nature, resulting from the Gaussian tail of measurement errors.

Figs.~\ref{xtlag} and \ref{xtlagf} demonstrate that our results are in good consistent with the three spectroscopic surveys. When combining all three surveys, the typical median and dispersion values are $\mu = -19$\,K and $\sigma = 197$\,K for \teff, and $\mu = -0.02$\,dex and $\sigma = 0.31$\,dex for \feh. In Fig.~\ref{xtlagqm}, we present the median differences across various stellar parameter ranges when comparing our results with the spectroscopic surveys. For \teff, there are no significant trends in the median differences ($\mu\Delta$\teff) across different \teff\ ranges. Our results align closely with those of Z23 and H22 and show slightly better consistency than Gaia23. For \feh, our results maintain a similar level of agreement with Z23 and H22 when \feh\ $> -1$\,dex. However, for \feh\ $< -2$\,dex, larger discrepancies are observed compared to the spectroscopic surveys.

\begin{figure*}
    \centering
    \includegraphics[width=0.5\textwidth]{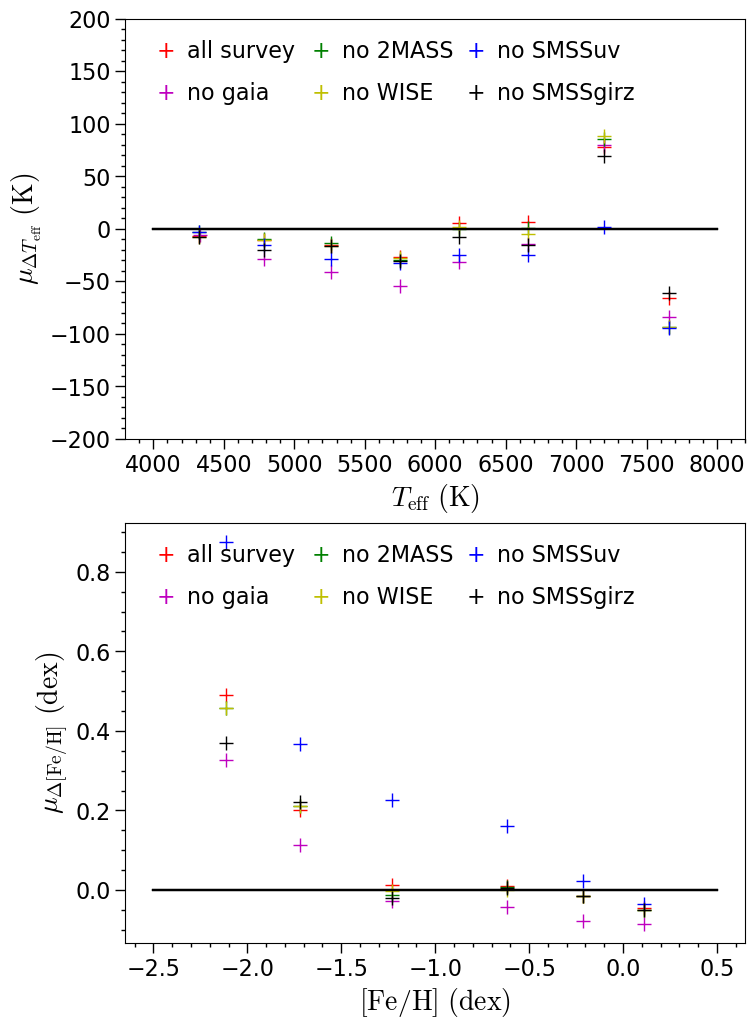}
    \caption{Median differences between the derived results in this study and those from the spectroscopic surveys for \teff\ (top panel) and \feh\ (bottom panel) when specific input survey data are missing. Different colors indicate cases where particular surveys are excluded.}
    \label{xtlaggm1}
\end{figure*}

When all measurements from the input catalogs are available (n=15), our results show improved agreement with the spectroscopic surveys, with differences of $\mu = -13$\,K and $\sigma = 176$\,K for \teff, and $\mu = -0.02$\,dex and $\sigma = 0.25$\,dex for \feh. However, not all stars have complete measurements across all input catalogs. To evaluate the effect of missing data, we randomly selected 10,000 sources with complete measurements (n=15) and tested how the absence of certain input catalogs impacts the derived parameters. Fig.~\ref{xtlaggm1} illustrates the median differences between our results and the spectroscopic surveys when specific input catalogs are excluded. 

For \teff, the deviations generally increase as \teff\ decreases, although the absence of any single input catalog does not substantially affect the overall deviations across the full \teff\ range. In contrast, for \feh, the deviations become more pronounced as \feh\ decreases. Notably, the absence of SMSS $uv$ data leads to significantly larger deviations in \feh, while the absence of other input catalogs has less impact on \feh\ deviations.

\begin{figure*}[htbp]
    \centering
    \includegraphics[width=0.8\textwidth]{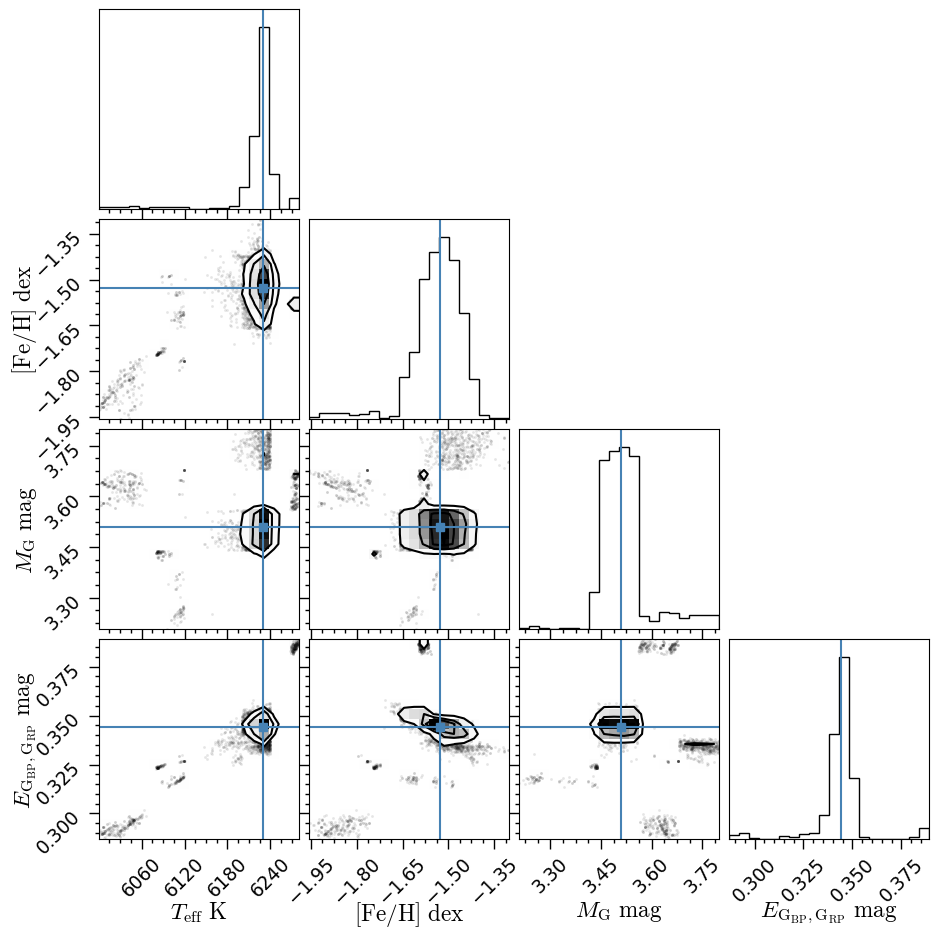}
    \caption{The corner plot presents the results of the MCMC fitting for an example source with the SMSS ID 62096272. The histograms depict the probability distributions of the model parameters. The contour lines illustrate the covariances between the parameters, indicating the degree to which they are correlated. The blue squares and lines denote the best-fit values for each parameter, providing a visual representation of the optimal parameter estimates.}
    \label{mct}
\end{figure*}

To investigate parameter correlations, we selected an example star (SMSS id = 62096272) and performed a MCMC analysis using 32 walkers and 5000 steps. Fig.~\ref{mct} shows the corner plot with one- and two-dimensional projections of the posterior distributions for the parameters. The contours illustrate covariances, while the histograms display Gaussian-like distributions. We observe that variations in effective temperature ($T_{\rm eff}$) significantly affect parameters such as extinction (\ebr) and absolute magnitude (\mg), but have a smaller impact on metallicity ([Fe/H]). We also note a minor degeneracy between extinction and metallicity, as metallicity can influence stellar color.

\subsection{Discussion of \ebr\ and $d$}

\begin{figure*}
    \centering
    \includegraphics[width=1\textwidth]{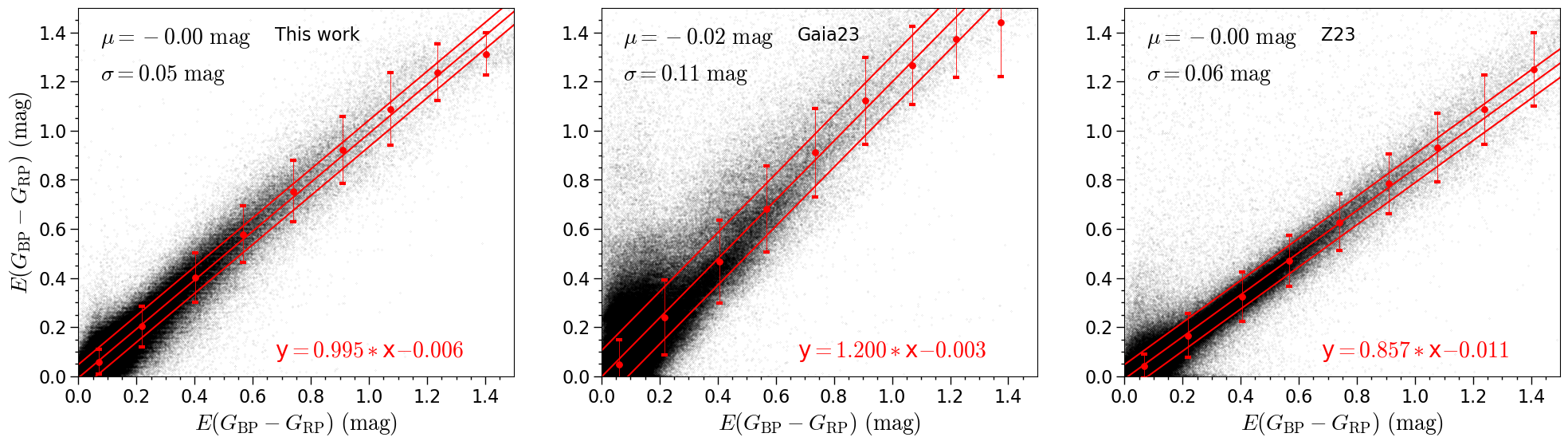}
    \caption{Comparison of \ebr\ values from this study, Gaia23, and Z23 (y-axis) with those derived from spectroscopic surveys (x-axis). The fit line equation is shown in the bottom right, and the median and standard deviation of the residuals are displayed in the top left of each panel. Red dots with error bars represent the mean and dispersion values of the data, while the red lines indicate the best-fit line and the lines offset by the dispersion.}
    \label{ebrspq}
\end{figure*}

Fig.~\ref{ebrspq} presents a comparison of the derived \ebr\ values from this study with those obtained using the star-pair algorithm based on spectroscopic survey parameters \citep{2023AJ....166..126S}. For reference, we also include the comparisons using the \ebr\ values from Gaia23 and Z23. All three datasets show good agreement with results from the spectroscopic data, though differences in the extinction laws adopted lead to variations in the slopes of the comparisons. Specifically, the slope for our results compared to the star-pair method is 1.0, while it is 1.2 for Gaia23 and 0.86 for Z23. In terms of residual dispersion, our results (0.05\,mag) are comparable to those of Z23 (0.06\,mag), whereas Gaia23 shows a larger residual dispersion of 0.11\,mag.

\begin{figure*}
    \centering
    \includegraphics[width=1\textwidth]{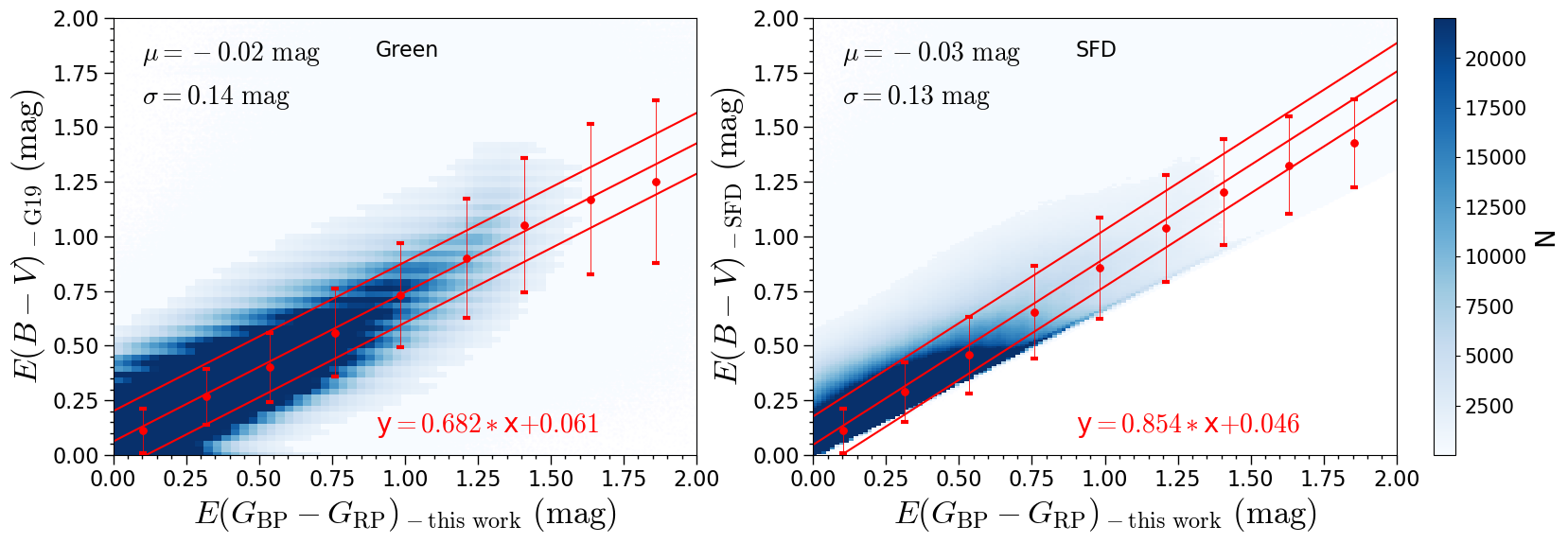}
    \caption{Comparison of \ebr\ values from this work (x-axis) with \ebv\ values from \citet{Green2019} (G19) and \citet{1998ApJ...500..525S} (SFD) (y-axis). The fit line equation is shown in the bottom right, and the median and standard deviation of the residuals are displayed in the top left of each panel. Red dots with error bars represent the mean and dispersion values of the data, while the red lines indicate the best-fit line and the lines offset by the dispersion.}
    \label{ebrsur}
\end{figure*}

We further compare our \ebr\ values with the \ebv\ estimates from G19 \citep{Green2019} and SFD98 \citep{1998ApJ...500..525S}, as shown in Fig.~\ref{ebrsur}. Strong linear correlations are observed, with residual variances of 0.14\,mag for G19 and 0.13\,mag for SFD98. For the G19 comparison, a linear regression yields a color excess ratio of $E(B-V)/E(G_{\rm BP}-G_{\rm RP}) \approx 0.68$, consistent with previous studies \citep{Chen2019, Sun2021}. When comparing with SFD98, we limit the sample to stars located more than 200\,pc above the Galactic plane. In this case, the color excess ratio is approximately 0.85, slightly higher than the G19 value. This difference may be attributed to SFD98's known 14\% overestimate of \ebv\ \citep{2010ApJ...725.1175S,2013MNRAS.430.2188Y}.

\begin{figure*}
    \centering
    \includegraphics[width=0.6\textwidth]{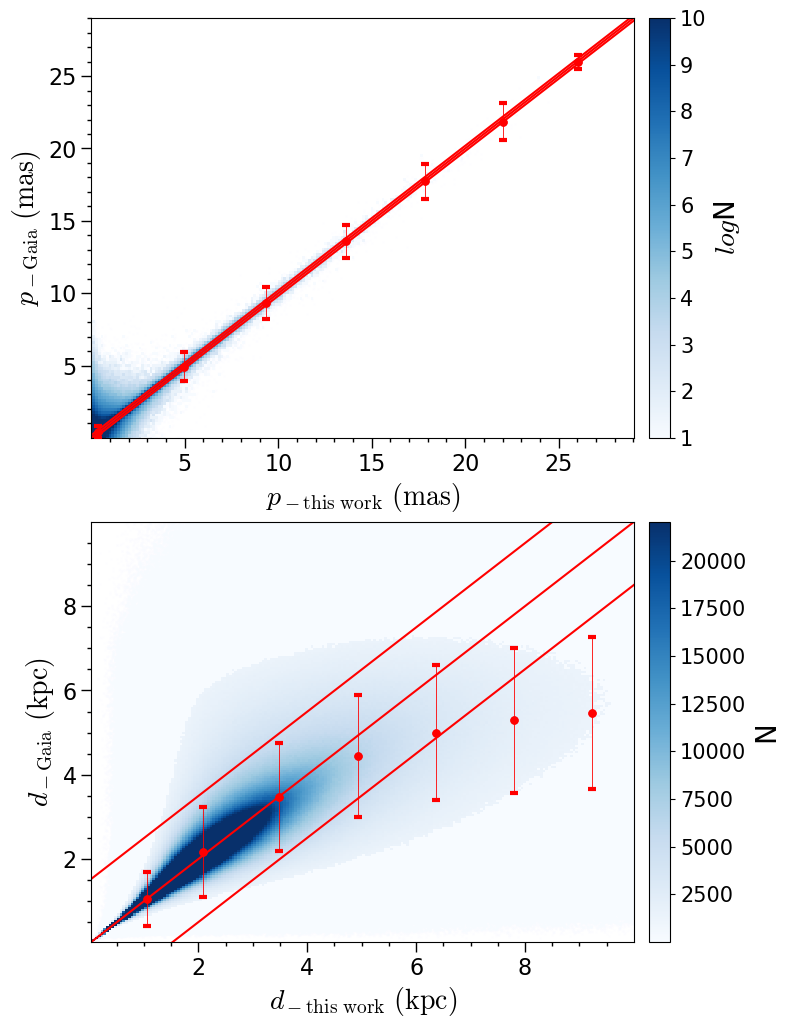}
    \caption{Comparison of our derived parallax and distance estimates with values from \citet{2023A&A...674A...1G}. The upper panel shows parallax, while the bottom panel shows distance. Red dots with error bars represent the mean and the dispersion values of the data, while the red lines indicate the line of equality and the lines offset by the dispersion.}
    \label{dfx}
\end{figure*}

To evaluate the accuracy of our distance estimates ($d$), Fig.~\ref{dfx} compares our results with parallax and distance measurements from Gaia DR3 \citep{2021AJ....161..147B}. Of the total sample, 99.9\% (140,450,761 sources) have Gaia parallax measurements. For most stars, our distance estimates are in good agreement with Gaia's values, with approximately 62\% of the sources showing relative discrepancies of less than 30\%. However, at larger distances, our estimates tend to be slightly higher than those from Gaia, likely due to the increasing uncertainties in parallax measurements over greater distances.

\subsection{Flags}

Table~\ref{cat} provides an overview of the contents of the final catalog, including detailed descriptions of all fields. To ensure users can assess the reliability of our derived parameters, we have implemented a comprehensive flagging system. These flags indicate the quality and completeness of the input data and are essential for assessing the trustworthiness of our results. The key flags include \texttt{flg\_[Fe/H]} (indicating the availability of the high-quality SMSS $v$ band in the input catalogs), \texttt{n} (the number of input measurements, such as multi-band photometry and Gaia parallax), \texttt{Chi-square} (the minimal $\chi^2$ value), \texttt{bands} (the names of the photometric bands used), and \texttt{flg\_par} (indicating the inclusion of Gaia parallax).

For \texttt{flg\_[Fe/H]}, a value of 1 indicates the presence of the SMSS $v$ band with an error less than 0.1\,mag in the input catalogs, contributing to more reliable metallicity (\feh) measurements. If the $v$ band is absent or does not meet this quality criterion, \texttt{flg\_[Fe/H]} is set to 0. The flag \texttt{n} represents the total number of input measurements, with a maximum value of 15, which includes multi-band photometry and Gaia parallax. While a higher \texttt{n} typically suggests greater accuracy, the reliability of the results also depends on the type and quality of the observations.

\begin{table}
\centering
\caption{Description of the Final Sample Catalog}
\label{cat}
\begin{tabular}{lll}
\hline
Field & Description & Unit \\
\hline
SMSS\_ID & SMSS object ID & \dots \\
sourceid & Gaia DR3 source ID & \dots \\
R.A. & Right Ascension from SMSS DR2 (J2000) & degree \\
Decl. & Declination from SMSS DR2 (J2000) & degree \\
Teff & Effective temperature & K \\
sigma\_Teff & Uncertainty of effective temperature & K \\
\feh & Metallicity & dex \\
sigma\_\feh & Uncertainty of metallicity & dex \\
flg\_\feh & Indicates high-quality SMSS $v$ band in the input: '0' = absent, '1' = present & \dots \\
M\_G & Absolute magnitude in Gaia $G$-band & mag \\
sigma\_M\_G & Uncertainty of absolute magnitude in Gaia $G$-band & mag \\
E\_bprp & Color excess \ebr\ & mag \\
sigma\_E\_bprp & Uncertainty of color excess \ebr\ & mag \\
d & Distance & kpc \\
sigma\_d & Uncertainty of distance & kpc \\
l & Galactic longitude & degree \\
b & Galactic latitude & degree \\
X/Y/Z & 3D positions in the Galactic Cartesian system & kpc \\
sigma\_X/Y/Z & Uncertainties of 3D positions & kpc \\
parallax & Parallax from Gaia DR3 & mas \\
sigma\_parallax & Uncertainty of parallax from Gaia DR3 & mas \\
n & Number of adopted measurements & \dots \\
Chi-square & Minimal $\chi^2$ value & \dots \\
bands & Names of the adopted photometric bands & \dots \\
flg\_par & Indicates Gaia parallax in the input: '0' = absent, '1' = present & \dots \\
\hline
\end{tabular}
\end{table}

The \texttt{Chi-square} value, as described in Section~\ref{sec:emp} and \citet{2023AJ....166..126S}, represents the quality of the parameter fitting, with our catalog restricting $\chi^2$ to values below 10 to ensure reliability. The \texttt{bands} field lists the specific photometric bands used in the input catalogs. These include Gaia $G$, $G_{\rm BP}$, and $G_{\rm RP}$; 2MASS $J$, $H$, and $K_S$; WISE $W1$ and $W2$; and SMSS $uvgriz$. Finally, \texttt{flg\_par} indicates whether Gaia parallax measurements were included in the input, with a value of 1 for inclusion and 0 for exclusion. This provides users with an additional measure of the completeness of the input data used to derive stellar parameters.

\subsection{Galactic Metallicity Distribution}

\begin{figure*}[htbp]
\centering
\includegraphics[width=0.8\textwidth]{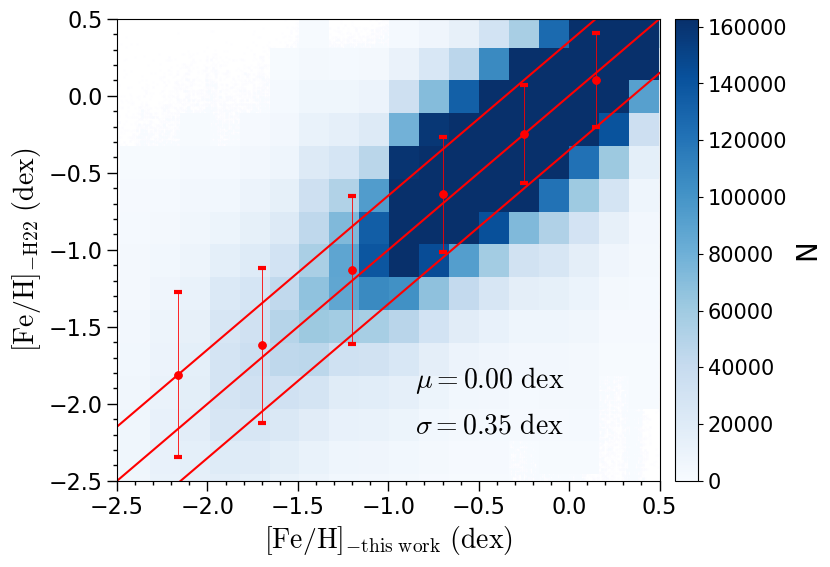}
\caption{Comparison of \feh\ values from this study (x-axis) with those from \citet{Huang2022} (H22) (y-axis). Red dots with error bars represent the mean and dispersion values. The red lines indicate the line of equality as well as the lines offset by the dispersion.}
\label{hfesur}
\end{figure*}

The inclusion of the SMSS $uv$ bands in the input star catalogs significantly improves the accuracy of \feh\ measurements. For this study, we focused on stars from the input catalogs that include the SMSS $v$ band with an error below 0.1 mag (indicated by \texttt{flg\_[Fe/H]} = 1) to analyze the metallicity distribution of the Milky Way. To validate our results, we compared the derived \feh\ values with those from \citet{Huang2022} (H22). As shown in Fig.~\ref{hfesur}, the \feh\ values from this study are in excellent agreement with those of H22, exhibiting a median difference of 0\,dex and a dispersion of 0.35\,dex. Importantly, our dataset includes \feh\ measurements for stars in low Galactic latitude regions, which were not covered in H22's work.

\begin{figure*}[htbp]
\centering
\includegraphics[width=0.65\textwidth]{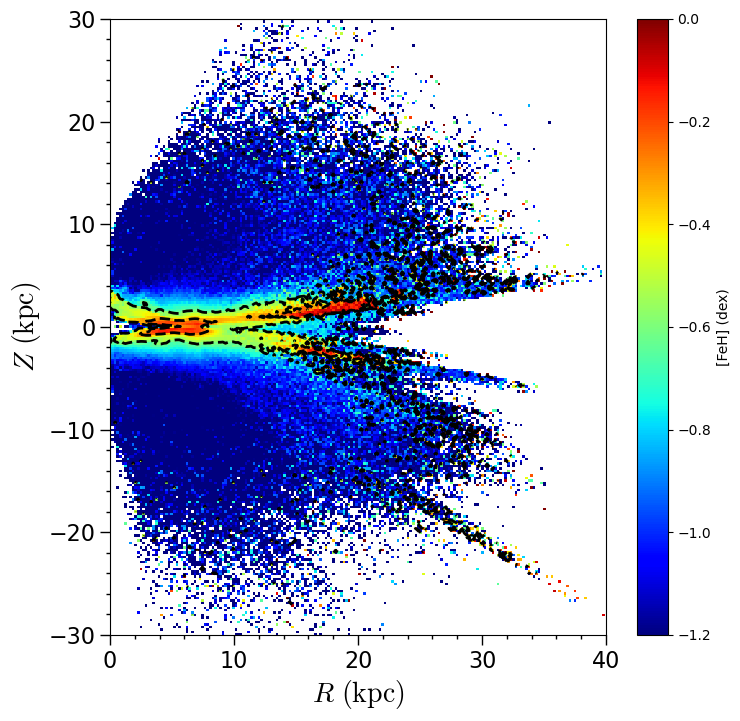}
\caption{Spatial distribution of median \feh\ values for stars meeting stringent criteria on distance and metallicity uncertainties, and with high-quality SMSS $v$-band detections, plotted in the $R$-$Z$ plane. Dashed lines represent contours of \feh.}
\label{rzfe1}
\end{figure*}

The spatial distribution of median metallicity (\feh) across the $R$-$Z$ plane is presented in Fig.~\ref{rzfe1}. For this analysis, we selected stars with high-quality \feh\ and distance measurements, applying stringent criteria: relative distance errors below 20\% ($\sigma_{d}/d < 0.2$), metallicity uncertainties less than 0.15 dex ($\sigma_{\rm [Fe/H]} < 0.15$ dex), and the presence of the SMSS $v$ band (\texttt{flg\_[Fe/H]} = 1). The figure highlights the Galactic metallicity gradient, with a metal-rich region concentrated in the Galactic disk at lower vertical distances ($Z$). As $Z$ increases, there is a clear transition to the metal-poor populations characteristic of the Galactic halo.

The metallicity distribution reveals radial variations within the Galactic disk. In the inner disk ($R < 8$ kpc), metallicity declines with increasing radius, while the outer disk displays significant flaring, characterized by metal-rich stars at higher vertical distances. However, near the midplane ($Z \sim 0$ kpc) at $R$ = 14–20 kpc, the average metallicity is not as high. This discrepancy is likely due to significant extinction, which reduces the number of observable stars and increases metallicity measurement uncertainties. Since our metallicity template library has an upper limit, metal-rich stars near this boundary may be misclassified as having lower metallicities. The smaller number of stars in this region further exacerbates this bias. A more detailed investigation of this phenomenon has not been undertaken in this paper due to its scope limitations. Nonetheless, the results presented here offer a comprehensive view of the Milky Way's stellar metallicity distribution, providing valuable insights into the Galaxy's chemical evolution. Our catalog serves as a robust foundation for future studies of the Galactic metallicity distribution function and related astrophysical topics. 

\subsection{The Dust Extinction Distribution in the Southern Sky}

\begin{figure*}[htbp]
\centering
\includegraphics[width=0.8\textwidth]{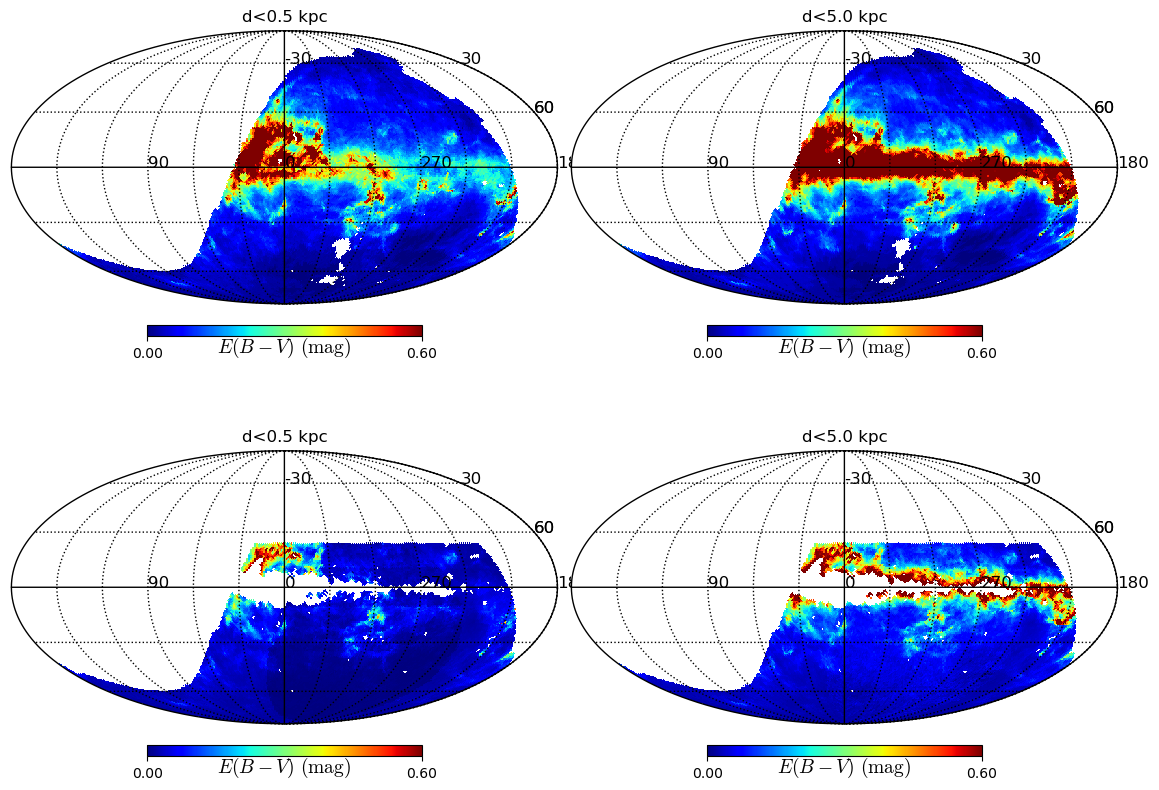}
\caption{Galactic dust extinction distributions within distances of $d < 0.5$ kpc (left) and $d < 5.0$ kpc (right) from this study (top) and \citet{Guo2021} (bottom), displayed in a Mollweide projection centered at $l = 0^{\circ}$. White areas indicate regions without data.}
\label{smap}
\end{figure*}

We present new Galactic extinction maps in the top panels of Fig.~\ref{smap}, constructed using the color excess \ebr\ derived in this study. To ensure high accuracy, these maps include stars with a color excess uncertainty of $\sigma_{{E(G_{BP}-G_{RP})}} < 0.06$ and a fractional distance uncertainty of $\sigma_{d}/d < 0.2$. 

\begin{figure*}[htbp]
\centering
\includegraphics[width=0.8\textwidth]{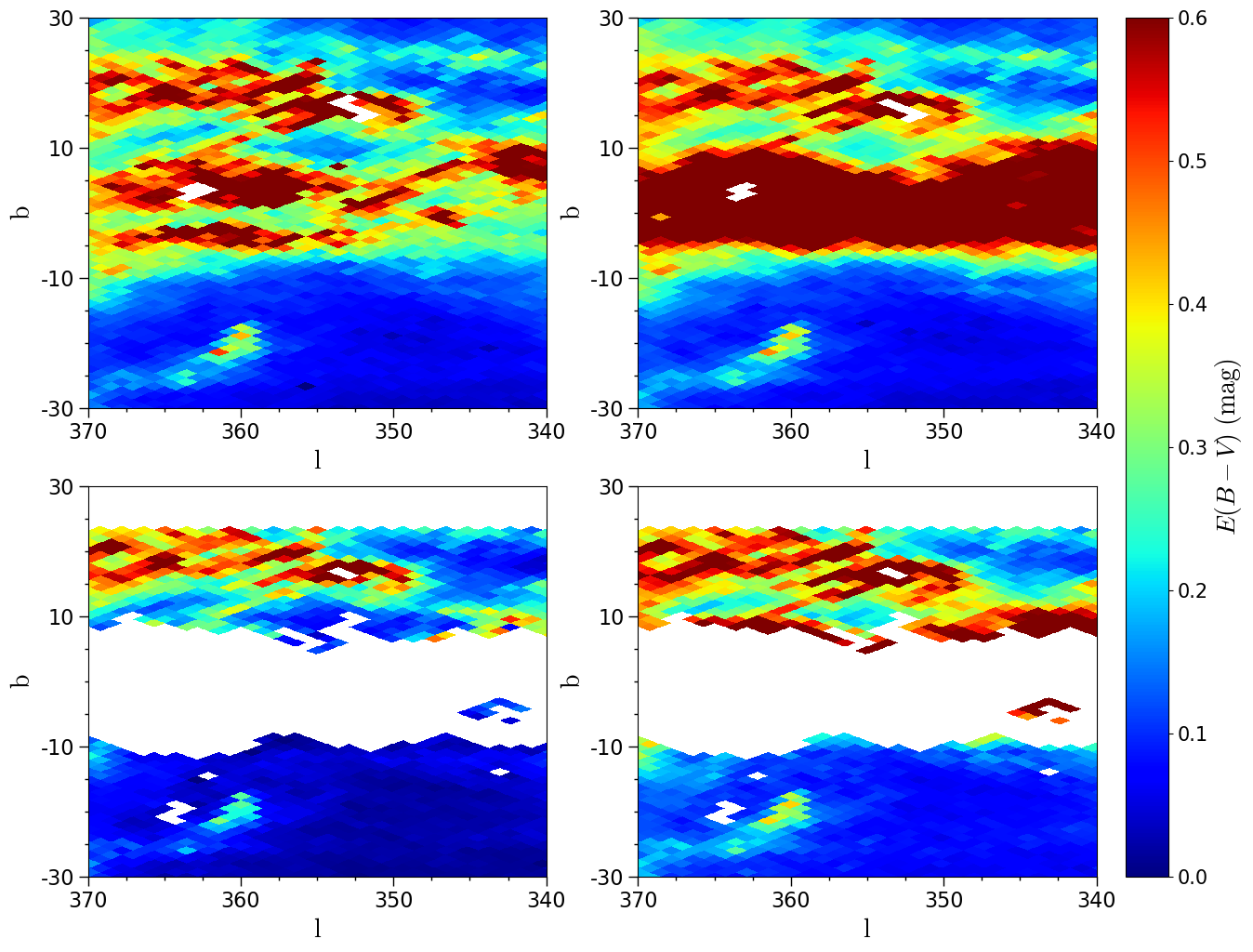}
\caption{Galactic dust extinction distributions within distances of $d < 0.5$ kpc (left) and $d < 5$ kpc (right) from this study (top) and \citet{Guo2021} (bottom) in the region $340^{\circ} < l < 370^{\circ}$ and $-30^{\circ} < b < 30^{\circ}$.}
\label{smapj}
\end{figure*}

Figures~\ref{smap} and \ref{smapj} compare the extinction maps from this work with those from \citet{Guo2021}. In Fig.~\ref{smap}, the extinction distribution for nearby stars within $d < 0.5$ kpc highlights the intricate structures of local molecular clouds. When extended to $d < 5$ kpc, the maps reveal the broader features of the Galactic dust disk. Covering an area of approximately 25,529 square degrees, these maps provide a detailed view of the dust distribution and confirm many of the structures reported by \citet{Guo2021}.

Figure~\ref{smapj} focuses on the region $340^\circ < l < 370^\circ$ and $-30^\circ < b < 30^\circ$, comparing the extinction maps over this specific area. The results from our study are consistent with those of \citet{Guo2021} in terms of the observed dust structures. However, our maps feature significantly more complete sky coverage, enabling a more comprehensive characterization of the Galactic disk. Additionally, the larger dataset used in this study allows for higher spatial resolution and greater depth, enhancing the reliability of the extinction measurements.

\section{Summary}\label{sec:sum}

In this study, we analyzed a sample of stars from the SMSS DR2 and applied the SPar algorithm to derive stellar parameters, including effective temperature (\teff), metallicity (\feh), absolute magnitude (\mg), extinction (\ebr), and distance ($d$) for 141 million stars. This work leverages the extensive multi-band photometric data from SMSS $uvgriz$, 2MASS $JH$\ks, WISE $W1W2$, and Gaia $G$\gbp\grp, complemented by Gaia parallaxes when available. The SPar algorithm employs a fitting procedure to match these observations against a stellar template library trained on data from LAMOST and Gaia, enabling precise determination of stellar physical parameters, extinction, and distances for individual stars. The resulting catalog is publicly accessible at \url{https://nadc.china-vo.org/res/r101372/}.

Our work represents a significant advancement in the characterization of stellar parameters for a vast region of the Galactic disk. Compared to previous studies within the SMSS framework, this study has significantly expanded both the sample size and the sky coverage. The derived stellar parameters, particularly \teff\ and \feh, exhibit excellent agreement with values from spectroscopic surveys, with deviations characterized by a dispersion of 195\,K for \teff\ and 0.31\,dex for \feh. These results are consistent with those reported in Z23 and H22, and show better precision than those from G23. Additionally, our estimates of reddening (\ebr) and distance ($d$) align well with prior studies, further validating the robustness of our approach.

The new dataset presented in this work offers a valuable resource for advancing our understanding of Galactic stellar populations, the interstellar dust distribution, and the chemical evolution of the Milky Way. Furthermore, the successful application of the SPar algorithm to such a large dataset demonstrates its potential for future large-scale surveys, such as Mephisto and the Chinese Space Station Telescope (CSST) optical survey. These surveys promise to provide even larger datasets, enabling the determination of stellar physical parameters, extinction, and distances on an unprecedented scale.

\begin{acknowledgements}
We would like to thank the referee for providing us with detailed and constructive feedback that has significantly enhanced the quality of the manuscript. This work is partially supported by the National Natural Science Foundation of China 12203016, 12173034, 12322304 and 12173013，Natural Science Foundation of Hebei Province No.~A2022205018, A2021205006, 226Z7604G, and Yunnan University grant No.~C619300A034, and Science Foundation of Hebei Normal University No.~L2022B33. W.Y.C. acknowledge the support form the science research grants from the China Manned Space Project. M.X.S. acknowledge the support of Physics Postdoctoral Research Station at Hebei Normal University. We acknowledge the science research grants from the China Manned Space Project with NO.~CMS-CSST-2021-A09, CMS-CSST-2021-A08 and CMS-CSST-2021-B03.

This research made use of the cross-match service provided by CDS, Strasbourg.

Guoshoujing Telescope (the Large Sky Area Multi-Object Fiber Spectroscopic Telescope LAMOST) is a National Major Scientific Project built by the Chinese Academy of Sciences. Funding for the project has been provided by the National Development and Reform Commission. LAMOST is operated and managed by the National Astronomical Observatories, Chinese Academy of Sciences.

This work presents results from the European Space Agency (ESA) space mission Gaia. Gaia data are being processed by the Gaia Data Processing and Analysis Consortium (DPAC). Funding for the DPAC is provided by national institutions, in particular the institutions participating in the Gaia MultiLateral Agreement (MLA). The Gaia mission website is https://www.cosmos.esa.int/gaia. The Gaia archive website is https://archives.esac.esa.int/gaia.

The national facility capability for SkyMapper has been funded through ARC LIEF grant LE130100104 from the Australian Research Council, awarded to the University of Sydney, the Australian National University, Swinburne University of Technology, the University of Queensland, the University of Western Australia, the University of Melbourne, Curtin University of Technology, Monash University and the Australian Astronomical Observatory. SkyMapper is owned and operated by The Australian National University’s Research School of Astronomy and Astrophysics. The survey data were processed and provided by the SkyMapper Team at ANU. The SkyMapper node of the All-Sky Virtual Observatory (ASVO) is hosted at the National Computational Infrastructure (NCI). Development and support the SkyMapper node of the ASVO has been funded in part by Astronomy Australia Limited (AAL) and the Australian Government through the Commonwealth’s Education Investment Fund (EIF) and National Collaborative Research Infrastructure Strategy (NCRIS), particularly the National eResearch Collaboration Tools and Resources (NeCTAR) and the Australian National Data Service Projects (ANDS).

This publication makes use of data products from the Two Micron All Sky Survey, which is a joint project of the University of Massachusetts and the Infrared Processing and Analysis Center/California Institute of Technology, funded by the National Aeronautics and Space Administration and the National Science Foundation.

This publication makes use of data products from the Widefield Infrared Survey Explorer, which is a joint project of the University of California, Los Angeles, and the Jet Propulsion Laboratory/California Institute of Technology, funded by the National Aeronautics and Space Administration.
\end{acknowledgements}

\bibliographystyle{raa}
\bibliography{ms2024-0312}

\end{CJK*}
\end{document}